\documentclass{PoS}
\usepackage{wrapfig}
\usepackage{graphicx}
\usepackage{mathtools}
\newcommand{\degree}{\ensuremath{^\circ}}

\title{Investigating the effects of finite resolution on observed transverse Rotation Measure distributions.}

\ShortTitle{The effects of finite resolution on RM gradients}

\author{\speaker{Eoin Murphy}, Colm Coughlan, Denise Gabuzda \\
        Physics Department, University College Cork, Cork, Ireland\\
        E-mail: \email{egm@student.ucc.ie}}

\abstract{Both the emission properties and evolution of Active Galactic Nuclei (AGN) radio jets are dependent on the magnetic (B) fields that thread them. Faraday Rotation measurements are a very important way of investigating these B fields, and can provide information on the orientation and structure of the B field in the immediate vicinity of the jet; for example, a toroidal or helical B field component should give rise to a systematic gradient in the observed Faraday rotation across the jet, as well as characteristic intensity and polarization profiles. However, real observed radio images have finite resolution, usually expressed via convolution with a Gaussian beam whose size corresponds to the central lobe of the point source response function. This will tend to blur the transverse structure of the jet, raising the question of how well resolved a jet must be in the transverse direction in order to reliably detect transverse structure. We  present the results of simulated Faraday rotation images designed to directly investigate the effect of finite resolution on observed transverse Faraday rotation measure structures.}

\FullConference{11th European VLBI Network Symposium \& Users Meeting,\\
		October 9-12, 2012\\
		Bordeaux, France}

\begin{document}

\section{Introduction}
In the standard theoretical model, the jets of AGN are electromagnetically launched and carry helical fields, which come about due to the rotation of the accretion disk plus the jet outflow. One of the primary observations that can support the presence of helical fields in the immediate vicinity of VLBI jets is the detection of transverse Faraday rotation measure (RM) gradients. 
\newline
Faraday Rotation occurs when an electromagnetic wave propagates through a  region with  plasma and a magnetic field. Faraday Rotation rotates the polarization of the electromagnetic wave because the left circularly polarized component of the EM wave has a different refractive index than the right circularly polarized component. The amount of rotation is  given by
\begin{equation}
\chi -\chi_0 = RM \lambda^2 
\end{equation}
where $\chi$ is the observed polarization angle, $\chi_0$ is the emitted polarization angle, $\lambda$ is the wavelength and RM is the Rotation Measure;
\begin{equation}
RM=\frac{e^3}{8 \pi^2 \epsilon_{0} m_{e}^2 c^3} \int n_e \overrightarrow{B} \bullet \overrightarrow{dl}
\end{equation}
where e is the elementary charge, $\epsilon_0$ is the permittivity of free space, $m_{e}$ is the mass of the electron, $n_e$ is the number density of electrons in the plasma and B is the magnetic field strength in the plasma. We would expect to observe a transverse RM gradient if a helical B field threads the jet, due to the systematic change in $\overrightarrow{B}$ $\bullet$ $\overrightarrow{dl}$ across the jet \cite{Blandford}.
\newline

These transverse RM gradients will be affected by the resolution of the observations used to detect them and so, in order to use these as identifiers for potential helical magnetic structure in AGN, it is very important to understand the effects of finite resolution on observed transverse RM profiles. Taylor and Zavala \cite{Taylor} proposed that transverse RM distributions must have width of at least 3 beam widths for an observed gradient to be reliable. Monte Carlo Simulations by Hovatta et al. \cite{Hovatta} demonstrated a lack of spurious gradients in RM distributions spanning 1.5 - 2 beams, if the RM differences  were > 3$\sigma$. 
\newline

In this paper we investigate transverse resolution requirements using Monte Carlo simulations of model jets, of various widths, with transverse RM gradients.

\section{Analysis Procedure}

We  constructed a model source which had a transverse RM gradient across its jet, and carried out Monte Carlo simulations based on this model source. A simple, cylindrical model source which fell off in total intensity linearly along the jet axis was used. This resulted in a ``core-jet'' like model with a jet length of about 20 mas. The RM gradient was applied to the last three quarters of the source and the resultant I, Q and U maps were generated.
\newline

\begin{wrapfigure}[19]{r}{60mm}
\includegraphics[width=.5\textwidth]{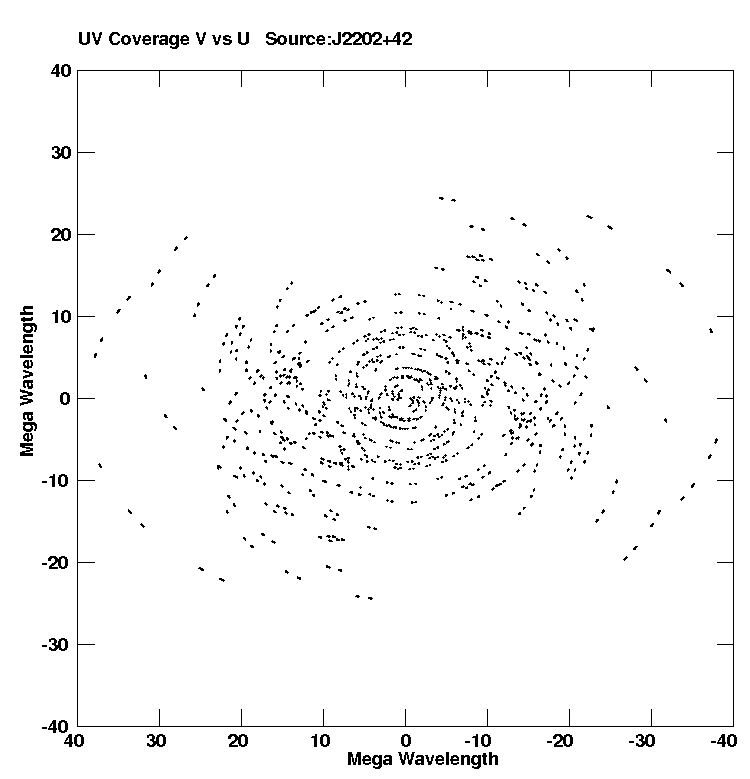}
\caption{The 22cm UV distribution, one of the UV distributions used during the analysis procedure.}
\label{UV}
\end{wrapfigure}

The direct Fourier transform of these model maps at UV point locations based on existing VLBA spanshot observations \cite{Coughlan} were then calculated in order to produce model visibility data with realistic UV coverage (Figure \ref{UV}). These data were generated for each of  four wavelengths, 18cm, 20cm, 21cm and 22cm. Random thermal noise and EVPA (Electric Vector Position Angle) calibration uncertainties of up to 3$\degree$ were added to these model visibilities. 
\newline

The stokes I, Q and U visibilites were then imaged in both CASA and AIPS using the natural-weight beam for the 22cm UV coverage. The Q and U images were then used to construct the corresponding polarization angle (PANG) images at each frequency, which were, in turn, used to construct RM images in the usual way.  Monte Carlo RM maps were constructed, based on 20 independent realizations of the thermal noise and EVPA calibration uncertainty, and an average RM map was derived by averaging together all 20 individual realizations of the RM distribution.
\newline

This entire procedure was carried out for model sources with intrinsic jet widths of 0.1, 0.2, 0.4 and 0.8 beam widths. The intrinsic RM gradient across the jet extends from + 30 rad m~$^{-2}$ to -30 rad m$^{-2}$, while the RM near the peak of the map was zero. This allowed for a direct investigation of the effects of finite resolution on RM gradients. 
\newline

\section{Results}

\begin{figure}
\includegraphics[width=\textwidth]{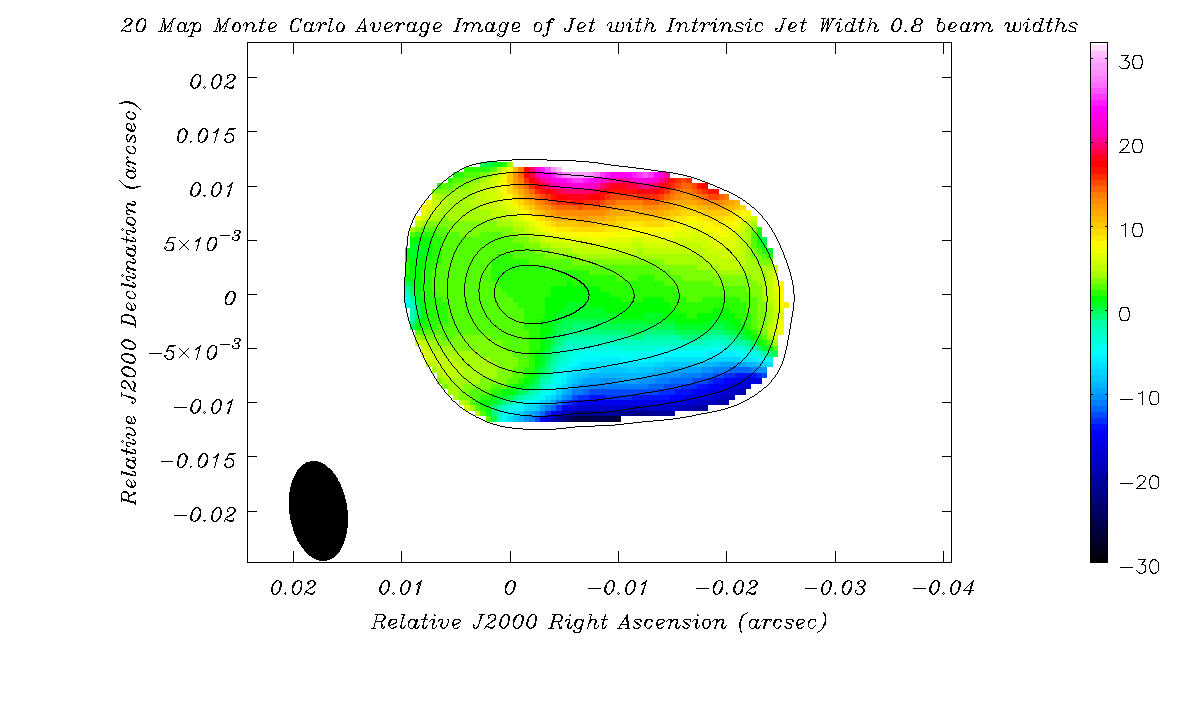}
\caption{Average RM map of intrinsic jet width  0.8 beam widths. This model map has a total flux of 2.4 Jy, a peak flux of 744 mJy/beam and an RMS of 0.5 mJy/beam.The contours are 1\%, 2\%,5\%, 10\%, 20\%, 40\% and 80\% of the peak flux. The error cutoff point for the RM map is 10 rad m$^{-2}$.}
\label{EVN_80}
\includegraphics[width=\textwidth]{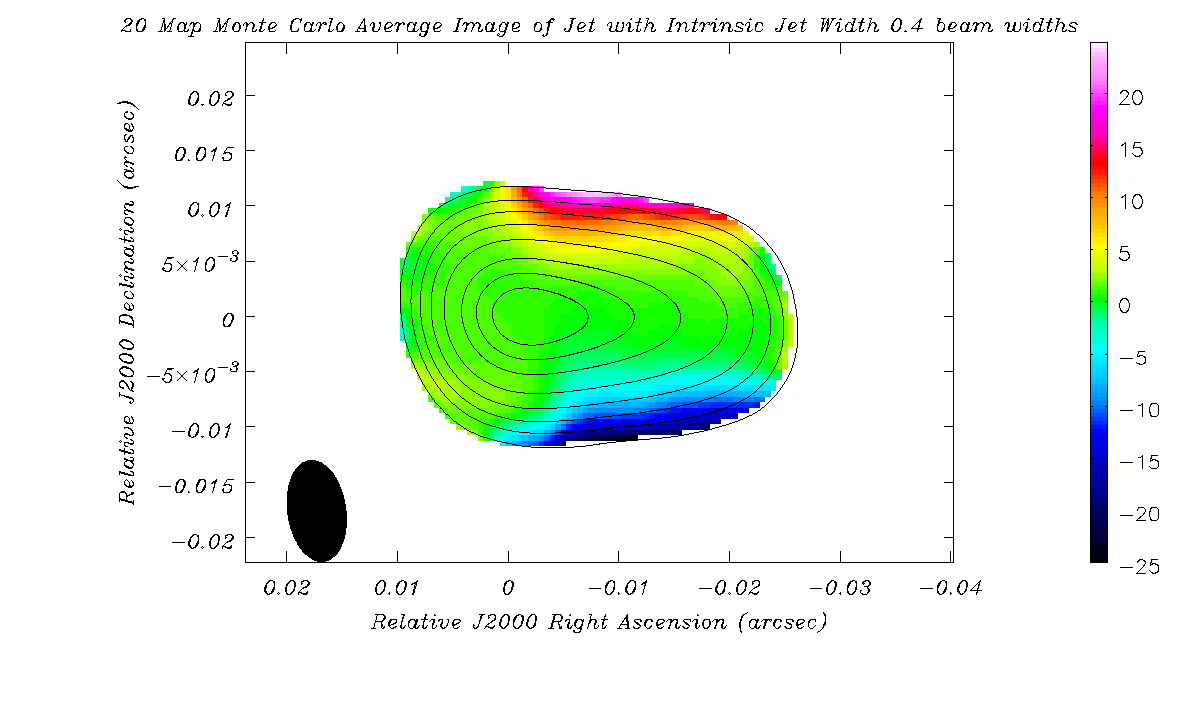}
\caption{Average RM map of intrinsic jet width  0.4 beam widths. This model map has a total flux of 2.4 Jy, a peak flux of 772 mJy/beam and an RMS of 0.55 mJy/beam.The contours are 1\%, 2\%,5\%, 10\%, 20\%, 40\% and 80\% of the peak flux. The error cutoff point for the RM map is 10 rad m$^{-2}$.}
\label{EVN_40}
\end{figure}

\begin{figure}
\centering
\includegraphics[width=\textwidth]{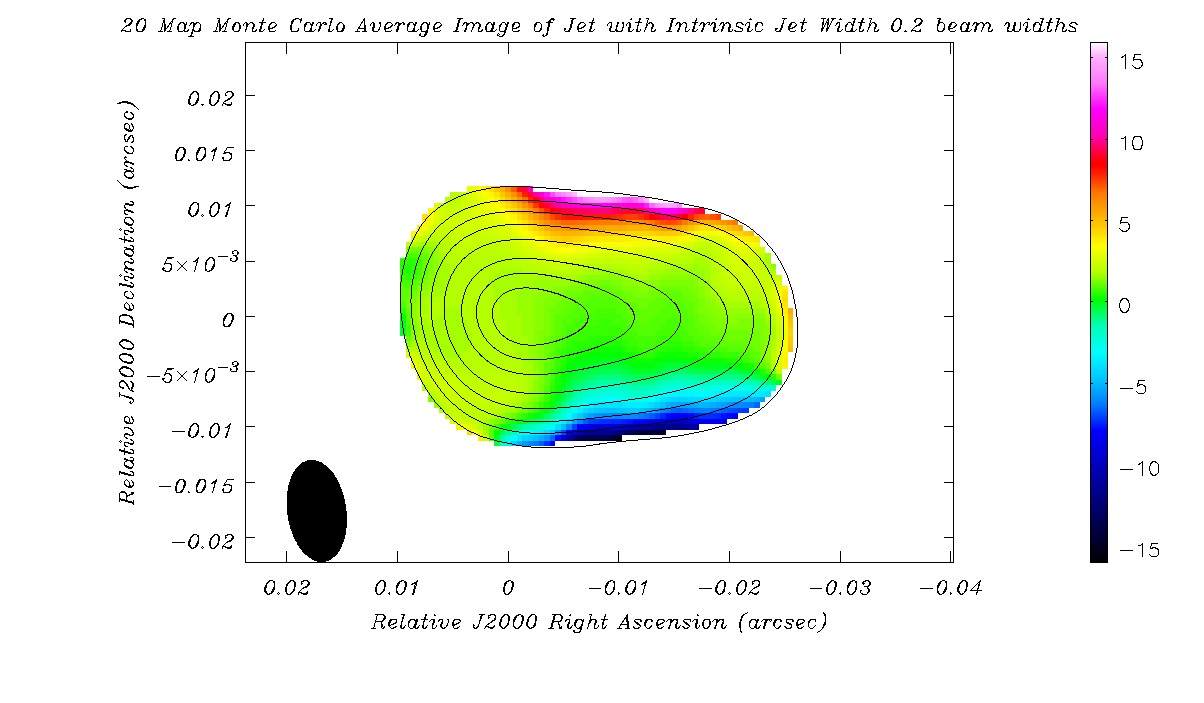}
\caption{Average RM map of intrinsic jet width  0.2 beam widths. This model map has a total flux of 2.4 Jy, a peak flux of 774 mJy/beam and an RMS of 0.45 mJy/beam. The contours are 1\%, 2\%,5\%, 10\%, 20\%, 40\% and 80\% of the peak flux. The error cutoff point for the RM map is 10 rad m$^{-2}$.}
\label{EVN_20}
\includegraphics[width=\textwidth]{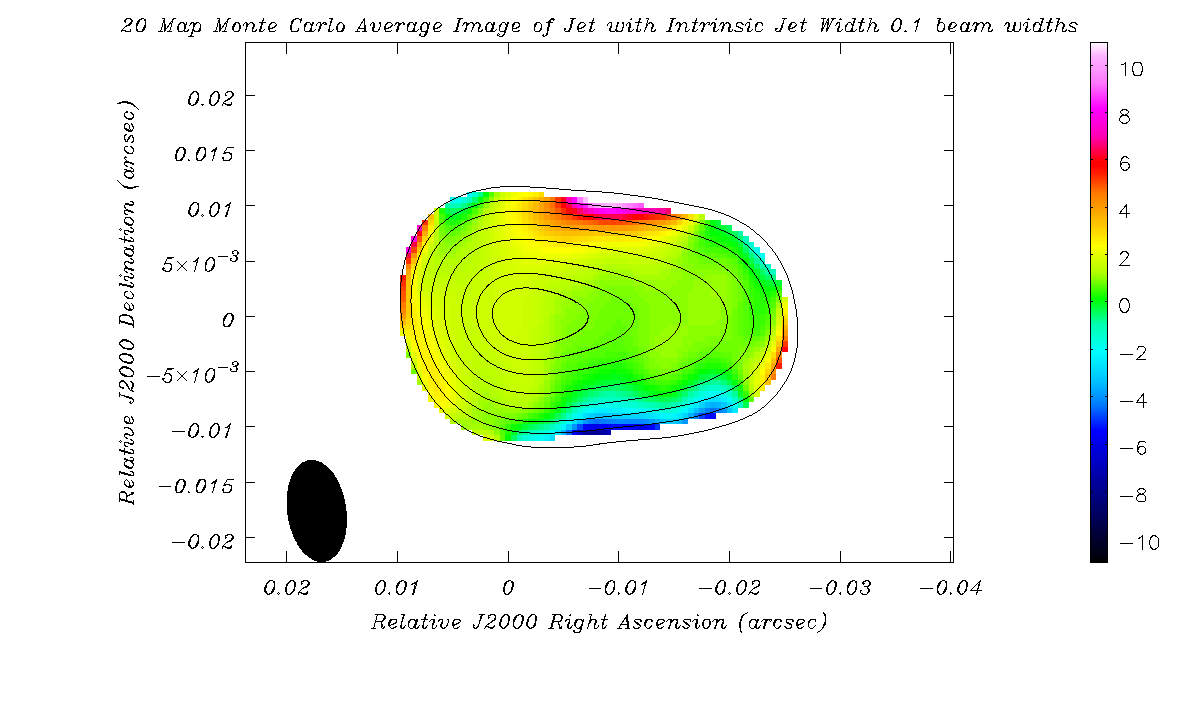}
\caption{Average RM map of intrinsic jet width  0.1 beam widths. This model map has a total flux of 2.4 Jy, a peak flux of 782 mJy/beam and an RMS of 0.5 mJy/beam. The contours are 1\%, 2\%,5\%, 10\%, 20\%, 40\% and 80\% of the peak flux. The error cutoff point for the RM map is 10 rad m$^{-2}$.}
\label{EVN_10}
\end{figure}

The Monte Carlo average RM simulations are shown in Figs. \ref{EVN_80} - \ref{EVN_10}. In all cases, the RM gradients that were introduced into the simulated data are visible in the ``noisy'' RM maps that were obtained, though the range of the RM gradients has been reduced significantly for the narrower jets. While individual RM maps for the narrower jets can be significantly distorted by noise, averaging together all the individual noisy  RM maps confirms the presence of the RM gradients. Individual RM maps with RM differences of over 3$\sigma$ were found for all 4 jet widths.\newline 

These results are broadly consistent  with Fig. 30 of Hovatta et al. \cite{Hovatta}, which shows that the fraction of ``false positives'' that were obtained in their Monte Carlo simulations did not exceed $\sim$1\% when a 3$\sigma$ criterion was imposed for the RM gradient, even when the observed width of the RM gradient was 1.5 beam widths (in contrast to the proposed criterion of Taylor and Zavala \cite{Taylor} of 3 beam widths). This suggests that signal to noise ratio, rather than resolution, provides the ultimate limitation on the visibility of RM gradients.

\section{Conclusions}
We have carried out Monte Carlo simulations of cylindrical jet-like structures with transverse RM gradients to investigate the resolution required to detect these gradients. These simulations show that 3$\sigma$ RM gradients remain clearly visible for intrinsic jet widths substantially smaller than the  beam width, though the actual RM values are suppressed by convolution. These results are broadly consistent with the results of Hovatta et al. \cite{Hovatta}. 
\newline

The three beam width criterion of Taylor and Zavala \cite{Taylor} was intended to ensure that observed transverse RM gradients were sufficiently well resolved. However, our demonstration that transverse RM gradients can be detected even when the transverse structure is poorly resolved (intrinsic jet width much less than a beam width) shows that such a criteria is not justified or meaningful. The proposed three beam width criterion is, therefore, too severe. Observed RM gradients should not be dismissed based on their observed width. Instead, their reliability should be estimated based on the RM difference across the jet ($\geq $ 3$\sigma$), the extent of the RM gradient along the jet, the quality of the RM fits and the possibility of distortion by optical depth effects.
\newline

Future work will include further Monte Carlo simulations for a variety of jet widths, UV coverages and wavelength ranges, to better determine limits to observing transverse structures under particular conditions.

\section{Acknowledgements}
Funding for this research was provided by the Irish Research Council for Science Engineering and Technology (IRCSET). Radionet 3 provided financial support to attend the 11$^{th}$ EVN Symposium and present this paper.


\begin{thebibliography}{4}

\bibitem{Blandford}Blandford R. D.,  \emph{Astrophysical Jets, Cambridge University Press 1993}, p. 26

\bibitem{Taylor}Taylor, G B, and R Zavala., 2010,  \emph{Are There Rotation Measure Gradients Across AGN Jets?}, \emph{ApJ (Letters)}, Volume 722, Issue 2,  L183-L187

\bibitem{Hovatta}Hovatta T., Lister M.L., Aller M.F., Aller H.D., Homan D.C., Kovalev Y., Pushkarev A.B., Savolainen T., 2012, \emph{MOJAVE: Monitoring of jets in Active Galactic Nuclei with VLBA experiments}, \emph{AJ} 144,105

 \bibitem{Coughlan} Coughlan, C.P., Murphy, R., Mc Enery, K., Patrick, H., Hallahan, R., Gabuzda, D.C., 2010, \emph{Proc. $10^{th}$ EVN Symposium}, \pos{PoS(X EVN Symposium)046}.




\end{thebibliography}
\end{document}